\documentclass[review]{elsarticle}

\usepackage{amssymb}
\usepackage[utf8]{inputenc}
\usepackage{natbib}
\usepackage{graphicx}
\usepackage{caption}
\usepackage{subcaption}
\usepackage{tabularx}
\usepackage{color}
\usepackage{cite}
\usepackage{lineno}
\usepackage{hyperref}
\usepackage{multirow}
\usepackage[final]{pdfpages}
\usepackage{fancyvrb}
\usepackage{amsmath,amsfonts,amsthm,amssymb}
\usepackage{setspace}
\usepackage{Tabbing}

\usepackage{fancyhdr}
\usepackage{lastpage}
\usepackage{extramarks}
\usepackage{chngpage}
\usepackage{soul,color}
\usepackage{graphicx,float,wrapfig}
\usepackage{framed}
\usepackage{color}
\usepackage{epstopdf}
\usepackage{capt-of}
\usepackage{mathrsfs} 
\usepackage{graphicx}
\usepackage{braket}
\usepackage{subcaption}
\usepackage[labelformat=parens,labelsep=quad,skip=3pt]{caption}

\setcitestyle{square}

\journal{PRE}

\begin{document}
\begin{frontmatter}


\author{Chinmoy Samanta}
\ead{samantachinmoy111@gmail.com}
\author{Aniruddha Chakraborty}
 

\title{Reaction-diffusion dynamics through a Gaussian sink in the presence of an attractive step wise linear potential energy curve}

\address{School of Basic Sciences, Indian Institute of Technology Mandi,
Kamand, Himachal Pradesh, 175005, India }

\begin{abstract}
    \noindent In the present report, we have introduced the Fredholm integral method to solve the Smoluchowski equation in the Laplace domain. We get an exact semi analytical solution for the linear potential energy curve in the dynamic diffusion process, and the survival probability is calculated by the numerical inverse Laplace transform method. We apply our method in two different physical context for finding different observable like average rate constant in electronic relaxation in solution and quantum yields in a photosynthetic systems or doped molecular crystal. 
\end{abstract}
\end{frontmatter}
\section{Introduction}
The dynamics of a random walker in the potential field has a significant contribution in a reaction-diffusion system. Not only in the reaction-diffusion system, but also it is quite omnipresent that the motion of a particle in nature is random, and the particle executes random walk. We have interested the cases where the motion is bounded in a given region space. The relevant situation could be found in a variety of macroscopic and microscopic systems and phenomena, including home range formation in animal behavior,  flocking, enzyme ligand binding, and artificial photosynthetic machines. \citep{wade1998electrostatic,livesay2003conservation,aldana2007phase,giuggioli2006theory}. Many problems of interest, viz. proton \citep{samanta1990deuterium} or electron-transfer processes \citep{marcus1985electron} involving an activated barrier crossing or barrierless processes such as relaxation from an excited state \citep{bagchi1990dynamics,bagchi1983theory}, etc. The general study involves the assumption that in the absence of the reaction phenomenon the motion of the walker is considered to be translationally invariant \citep{rice1985comprehensive,wilemski1973general,kenkre1983exciton,szabo1984localized,spendier2013analytic}. In this type of systems, all that is relevant is the initial distance between the particle and reactive site, the manner of motion that occurs in between, and the rate of reaction. In the present paper, we extended these studies fundamentally by going beyond that assumption. We study the system in which a particle diffuse in a piecewise linear potential, and it gets captured at a given rate when it passes through a Gaussian sink. The position of the sink is not necessarily to be at the center of the attractive potential, and it can be elsewhere on the potential energy curve also. It is also noticed that the shape of the sink function is usually assumed to be the Dirac delta function. Many articles are devoted to solving the Smoluchowski equation with the localized Dirac delta sink on various potential energy curves \citep{bagchi1983theory,sebastian1992theory,spendier2013reaction,chase2016analysis,saravanan2019reaction} in time and Laplace domain. The main advantage of using this localized sink is that we may get an exact analytical solution of the equation for some known potential. The choice of delta function sink makes the model successful for explaining the decay processes, specifically in the barrierless electronic relaxation process. Although the most acceptable sink function could be a delocalized function and because it makes the function more generalized than the localized one. As a result, the analytical solution of the equation is not found yet even for simple flat potential. This motivates us to investigate with the Gaussian sink function for different potential energy curve.\\
In this paper, we present a semi-analytical method for solving the Smoluchowski equation for the Gaussian sink on the linear potential energy curve. We derive the solution in the Laplace domain and the time domain survival probability is calculated by numerical inversion of the Laplace domain solution. Our method involves the solution of the Fredholm integral equation of the second kind.
\section{Methodology}
For the above model of reaction dynamics, the time evaluation equation for the probability distribution of finding the particle at position x at time t before the reaction is often described by the Smoluchowski equation, given that initially the particle at $x_{0}$,
\begin{equation}
      \frac{\partial P(x,t|x_{0})}{\partial t}=[\mathscr{L}-S(x)-k_{r}]P(x,t|x_{0}),
\end{equation}
with 
\begin{equation}
    \mathscr{L}=D\frac{\partial^2}{\partial x^2}+ \frac{D}{k_{B}T}\frac{\partial}{\partial x}\frac{d}{d x}U(x).
\end{equation}
The above stochastic equation could be obtained in the usual manner \citep{reichl1999modern, risken1989}. The origin of the equation is the Langevin equation with white noise for the random walker, including the assumption that the particle executes random walk in a highly damped environment.
The boundary and initial conditions are given by
\begin{equation}
  \begin{split}
        \frac{\partial}{\partial x}P(x,t|x_{0})|_{x=x_{c}}=0, & \lim_{x\xrightarrow{}\infty}P(x,t|x_{0})=0,\\
         P(x,t=0|x_{0})= \delta(x-x_{0}).
  \end{split}
\end{equation}
In Eq. (1), Eq. (2) and Eq. (3), S(x) denotes the sink function, in the current study it is assumed to be a Gaussian function, with the center at $x_{c}$. D describes the diffusion constant, which is assumed to be constant in this study, and $k_{r}$ is the position independent radiation rate constant. $k_{B}$ and $T$ are Boltzmann's constant and absolute temperature. The excited state potential energy curve of the particle is given by $U(x)$.
In the Laplace domain, the Eq. (1) becomes
\begin{equation}
     [s+k_{r}-\mathscr{L}+S(x)]\mathscr{P}(x,s|x_{0})=P(x,t=0|x_{0}).
\end{equation}
We can get the Green's function for the linear potential, $U(x)=\beta|x|$, from the following equation
\begin{equation}
     [s+k_{r}-\mathscr{L}+S(x)]\mathscr{P}(x,s)=\delta (x-x_{0}).
\end{equation}
 By using boundary conditions, the Green's function for the linear or 'V' shaped potential in the absence of any sink function is given by \citep{chase2016analysis}
\begin{equation}
    \mathscr{G}_{0}(x,s+k_{r}|x_{0})=\frac{e^{-(|x|-|x_{0}|)/2l}}{\Gamma \sqrt{1+\frac{4l(s+k_{r})}{\Gamma}}}[e^{-\sqrt{1+\frac{4l(s+k_{r})}{\Gamma}}|x-x_{0}|/2l}+ \frac{e^{-\sqrt{1+\frac{4l(s+k_{r})}{\Gamma}}(|x|+|x_{0}|)/2l}}{\sqrt{1+\frac{4l(s+k_{r})}{\Gamma}}-1}].
\end{equation}
Where $\Gamma=\beta D/k_{B}T$, it is proportional to the strength of the potential in velocity unit at the thermal equilibrium, and $l=D/\Gamma$, it corresponds to the characteristic width of the steady state distribution.
Green's function solution for an arbitrary shape of the sink function would be 
\begin{equation}
    \mathscr{G}(x,s|x_{0})=\mathscr{G}_{0}(x,s|x_{0})-k_{c}\int_{-\infty}^{\infty} dy \mathscr{G}_{0}(x,s|y) S(y) \mathscr{G}(y,s|x_{0}).
\end{equation}
Here $k_{c}$ is the strength of the sink and we consider  $S(x)= e^{-\frac{(x-x_{c})^2}{2 \sigma^2}}/\sqrt{2 \pi \sigma^2}$ in our present study. $\sigma$ describe the finite width of the sink function.
The survival probability in the Laplace domain is given by
\begin{equation}
    \hat{Q}(x_{0},s)=\int_{-\infty}^{\infty} dx \mathscr{G}(x,s|x_{0}).
\end{equation}
By using the detailed balance conditions, we can farther write the above equation as
\begin{equation}
    \mathscr{G}(x_{0},s|x)=\mathscr{G}_{0}(x_{0},s|x)-k_{c}\int_{-\infty}^{\infty} dy \mathscr{G}_{0}(y,s|x) S(y) \mathscr{G}(x_{0},s|y).
\end{equation}
Corresponding reaction probability would be
\begin{equation}
    \hat{R}(x,s)=s^{-1}-\hat{Q}(x,s)
\end{equation}
Then we can find the equation for the reaction probability as
\begin{equation}
    \hat{R}(x,s)=s^{-1}k_{c}\int_{-\infty}^{\infty} dy \mathscr{G}_{0}(y,s|x) S(y)-k_{c}\int_{-\infty}^{\infty} dy \mathscr{G}_{0}(y,s|x) S(y) \hat{R}(y,s).
\end{equation}
We can modify the form of the above equation into
\begin{equation}
    \begin{split}
    \hat{R}(x,s)=&s^{-1}k_{c}\int_{-\infty}^{\infty} dy \mathscr{G}_{0}(y,s|x)S(y)\\
    &-\hat{R}(x,s)k_{c}\int_{-\infty}^{\infty} dy \mathscr{G}_{0}(y,s|x) S(y) \hat{R}(y,s)\hat{R}(x,s)^{-1}.
    \end{split}
\end{equation}
Farther we can have
\begin{equation}
   \hat{R}(x,s)=\frac{k_{c}}{s}\frac{\int_{-\infty}^{\infty} dy \mathscr{G}_{0}(y,s|x)S(y)}{1+k_{c}\int_{-\infty}^{\infty} dy \mathscr{G}_{0}(y,s|x) S(y) \hat{R}(y,s)\hat{R}(x,s)^{-1}}
\end{equation}
Since the Eq. (11) got a Fredohlm integral form of first kind then the series solution of  would be \citep{polyanin1998handbook}
\begin{equation}
     \hat{R}(x,s)=\frac{k_{c}}{s}J_{1}(x,s)+\lambda \frac{k_{c}}{s}J_{2}(x,s)+\lambda^2\frac{k_{c}}{s}J_{3}(x,s)+......
\end{equation}
with $\lambda=-k_{c}$ and
\begin{equation}
    \begin{split}
       J_{1}(x,s)=&\int_{-\infty}^{\infty} dy \mathscr{G}_{0}(y,s|x)S(y),\\
       J_{2}(x,s)=&\int_{-\infty}^{\infty} dy_{1} \mathscr{G}_{0}(y_{1},s|x)S(y_{1})J_{1}(y_{1},s),\\
        & :\\
       J_{n}(x,s)=&\int_{-\infty}^{\infty} dy_{n-1} \mathscr{G}_{0}(y_{n-1},s|x)S(y_{n-1})J_{n-1}(y_{n-1},s)\\
       & :
     \end{split}
\end{equation}
We consider the term, $\frac{\hat{R}(y,s)}{\hat{R}(x,s)}$ in Eq. (12) as
\begin{equation}
   \begin{split}
       \frac{\hat{R}(y,s)}{\hat{R}(x,s)}&=
       \frac{J_{1}(y,s)+\lambda J_{2}(y,s)+\lambda^2J_{3}(y,s)+......}{J_{1}(x,s)+\lambda J_{2}(x,s)+\lambda^2J_{3}(x,s)+......}
   \end{split}
\end{equation}
The expression of $\hat{R}(x,s)$ in Eq. (13) now become
\begin{equation}
    \hat{R}(x,s)=\frac{1}{s}\frac{k_{c}J_{1}(x,s)}{1+k_{c}\sum_{i=2}^{\infty} \lambda^{i-2} J_{i}(x,s)/\sum_{i=1}^{\infty} \lambda^{i-1} J_{i}(x,s)}.
\end{equation}
In the above we get the expression for $\hat{R}(y,s)$ from  the solution of the Fredholm integral equation of the second kind. The expression in Eq. (17) gives the exact solution to the Eq. (11) for the finite width of a general delocalized sink function. One of the important consequences of this method is we can easily recover the exact solution even for the localized $\delta-$function sink from the above equation. By putting the sink function $S(x)=\delta(x-x_{c})$ into Eq. (17) we obtain
\begin{equation}
    \hat{Q}(s)=\frac{1}{s}[1-\frac{k_{c}\mathscr{G}_{0}(x_{c},s|x_{0})}{1+k_{c}\mathscr{G}_{0}(x_{c},s|x_{c})}].
\end{equation}
In this case all $J's$ expression reduced into 
\begin{equation}
    J_{n}(x,s)=\mathscr{G}_{0}(x_{c},s|x_{0})\mathscr{G}_{0}^{n-1}(x_{c},s|x_{c})
\end{equation}
It is noticeable easily that the series solution converges very fast, and the first term contributes the majority amount to the solution. Moreover, the series solution enters as a ratio in Eq. (13) and it does decrease the amount of error extensively for the series solution and gives rise almost exact solution. For the shake of convenience in the next calculations, we approximate the series solution up to the first term only and the Eq. (17) is simplified into  \begin{equation}
    \hat{R}_{a}(x,s)\cong\frac{1}{s}\frac{k_{c}J_{1}(x,s)}{1+k_{c}J_{2}(x,s)/J_{1}(x,s)}.
\end{equation}
The most desire result, survival probability in the Laplace domain for nonzero constant nonradiative decay rate $k_{r}$ can be found from Eq. (10) as
\begin{equation}
    \hat{Q}(s)=\frac{1}{s+k_{r}}[1-\frac{k_{c}J_{1}(x_{0},s)}{1+k_{c}J_{2}(x_{0},s)/J_{1}(x_{0},s)}].
\end{equation}
 Although we can make the solution complete by increasing the terms in series solution for the ratio, $ \frac{\hat{R}(y,s)}{\hat{R}(x,s)}$ in Eq. (14). For the present case of linear potential energy curve with the Gaussian sink function we see that the series solution converges very fast and for the convenient we consider the first term. The expression for average rate constant, $K_{I}$ of the decay process of the molecule could be found from
\begin{equation}
    K_{I}^{-1}=\lim_{s\xrightarrow{}0}\hat{Q}(s)
\end{equation}

\section{Result and Discussion}
\subsection{Electronic relaxation processes in polar solution}
\begin{figure}[hbt!]
\begin{subfigure}{6cm}
    \centering
    \includegraphics[width=5cm]{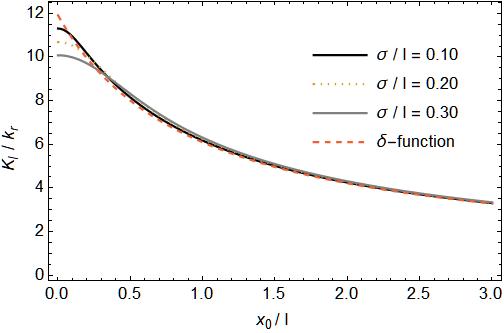}
    \caption{$x_{c}/l=0$}
\end{subfigure}    
\begin{subfigure}{6cm}
    \centering \includegraphics[width=5cm]{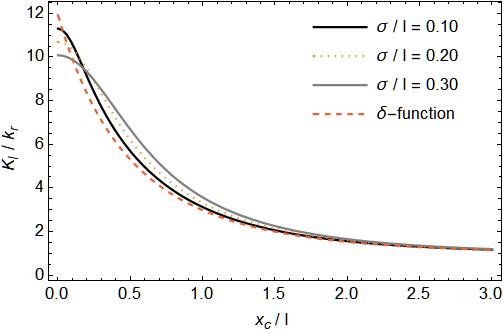}
    \caption{$x_{0}/l=0$}
\end{subfigure}    
 \caption{\small{\emph{Average rate $K_{I}$ (in units of $k_{r}$) of the relaxation process vs the location of particle $x_{0}$ (Figure (a)) as well as the location of the center $x_{c}$ (Figure (b)) of the Gaussian sink with keeping other fixed at the center of PEC. In both figures, each curve correspond to different width $\sigma$ including $\delta-$sink (dashed) with almost zero width. Other fixed dimensionless parameter values are $\Gamma/l k_{r} = 10$ and $l k_{r}/k_{c}=0.05$. A clear nonmonotonic dependence of $K_{I}$ on width $\sigma$ is observed when particle is little far away from the sink and we can also say from these figures that when the particle is far form the sink, rate does not depend upon the shape of the sink function.}}}
\end{figure}
In the following, we discuss a detailed analysis of the diffusion-reaction dynamics when a particle is an encounter by an attractive piecewise linear potential in the presence of Gaussian sink of finite width. We elaborate on the dependency of the average rate constant on different parameter as well as we do apply Numerical inverse Laplace teachings to visualize the survival probability of the particle during the relaxation processes. There are two common situations to consider with respect to the relative position of the particle and the sink. One is when the particle is initially located at the uphill position relative to the sink, and the other is when the particle is at the downhill location relative to the sink. To analysis the dynamics, we consider the downhill location to be at the center of the PEC, which is the origin. Since our solution is given in the Laplace domain, the most general observable to calculate is the average rate constant $K_{I}$. Figure 1, where $K_{I}$ are plotted against the position of the particle $x_{0}$ in Figure 1(a) and the center of the Gaussian sink $x_{c}$ in Figure 1(b) when $x_{c}$ and $x_{0}$ are fixed at the origin respectively. Each curve in these figure corresponding to different width $\sigma$ including $\delta-$sink. A nonmonotonic dependency of the rate concerning the width of the sink is observed. It is clear more in the case of the uphill location of the sink on the PEC. Also, we see an independency of the sink shape when the relative distance L between the particle and the sink is increasing. A direct measurement of $K_{I}$ with the variation of the width $\sigma$ can be seen in Figure 2. 
\begin{figure}[hbt!]
    \begin{subfigure}{6cm}
    \centering
   \includegraphics[width=5cm]{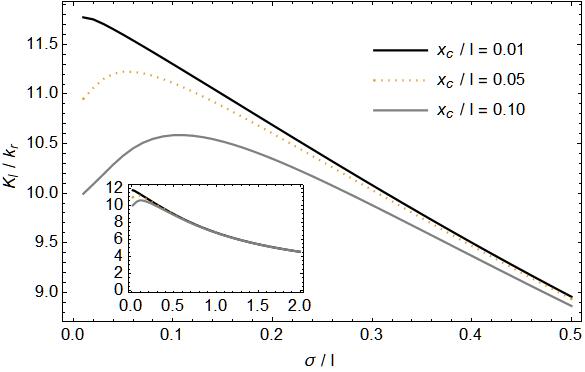}
    \caption{$x_{0}/l=0$ }
\end{subfigure}    
\begin{subfigure}{6cm}
    \centering \includegraphics[width=5cm]{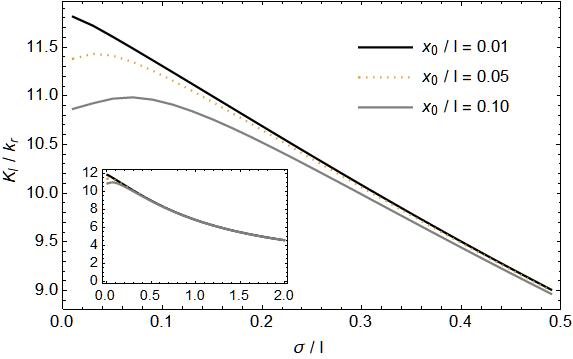}
    \caption{$x_{c}/l=0$ }
\end{subfigure}  
    \caption{\small{\emph{Average rate $K_{I}$ (in units of $k_{r}$) with the width $\sigma$ (in units of l) of the sink for different position of the centre of Gaussian sink $x_{c}$ with keeping the particle location $x_{0}$ at the center of PEC (Figure a). The opposite situation is shown in Figure (b). The rest of the parameter values are the same as in Figure 1. With the increase in $\sigma$, the same nonmonotonicity of $K_{I}$ is visible as in Figure 1.  }}}
\end{figure}
A little increase in $K_{I}$ is seen for short relative distance between the particle and the sink after that it decays monotonically. In the next, we calculate $K_{I}$ with the variation of the strength $\Gamma$ for the central placement of either the particle or the sink, when the other is varying symmetrically on both sides. Figure 3, all the left panels correspond to symmetrical movement of the particle for the centrally placed sink at the origin and the other situation are depicted in right panels where sink position is varying symmetrically around the particle at the origin. Top two panels describe $K_{I}$ for the $\delta-$ sink and rest of all are for Gaussian sink. In each case at the center of the PEC we get $K_{I}$ maxima. For the lower strength of the PEC we get the higher value of $K_{I}$. Here the cause could be the leading diffusive motion over the potential induced motion (drift) of the particle. An opposite characteristics is found i.e. higher strength $\Gamma$  gives higher value of $K_{I}$, when the particle gets separated by sufficient amount from sink. In this situation particle exhibits potential induced motion. Here we see both nonmonotonic nature of $K_{I}$ with respect to strength $\Gamma$ around the center as well as monotonic behavior far from the center. The noticeable thing is the length $l_{nmt}$ (say) over which this nonmonotonicity visible is almost same for fixed width of the sink width. Introduction of the sink influences length $l_{nmt}$, it is increased with width of the sink. \\
The most desirable physical entity in the diffusion dynamics is calculating the survival probability $Q(t)$. Most generally it is not possible to find the inverse Laplace transform of $\Tilde{Q}(s)$ when there is delocalized sink on some known potential energy curves including flat, linear and parabolic etc. A numerical Laplace inversion becomes necessary. But these transformation programs have the reputation of being inaccurate if the corresponding time-dependent function is oscillating. In the present case the probability function do not possess this feature. We calculate $\Tilde{Q}(s)$ numerically with the variation of different parameters, including strength $\Gamma$ and width $\sigma$, then find the time domain solution by using inverse Laplace transform tool in Mathematica software(v 11.3). Figure 4(a), when Q(t) is plotted against time t (in units of $x_{c}^{2}/2 D$) with variation of four different values of $\Gamma$ and the particle is placed at the center of the PEC and sink is placed at the uphill location. The opposite situation is shown in Figure 4(b). As it is expected Q(t) decays non monotonically with the strength $\Gamma$ for the uphill location of the sink and it decays monotonically when the particle is at uphill location relative to the sink. In these figures calculation are made by considering the finite width of the sink. So the presence of the attractive potential has both favorable and an unfavorable effect on decay processes. Figure 4(c) $\&$ 4(d), where Q(t) is plotted for the variation of the width $\sigma$ (in units of L) for two different configuration of particle and sink as in Figure 4(a) $\&$ 4(b) respectively. In both the cases width influenced decay of Q(t) is noticed. Thus the finite width of the sink function on the PEC has a significant role in the diffusion dynamics. All the findings above could be incorporated to various processes according to relative position of the particle and the sink. When the sink position is at the origin ($x_{c}$=0), i.e., the minimum of the potential well, corresponding to a barrierless process, encountered in the case of relaxation from an excited state \citep{bagchi1990dynamics,bagchi1983theory}. The case of initial position of the particle at the origin ($x_{0}$=0) is of importance in most of the activated barrier crossing type processes such as the one encountered in the case of electron transfer reactions \citep{marcus1985electron}.  

\begin{figure}[hbt!]
    \begin{subfigure}{6cm}
    \centering\includegraphics[width=5cm]{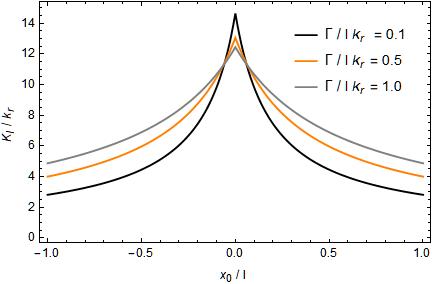}
    \caption{$x_{c}/l=0$, $\delta-$sink }
    \end{subfigure}
    \begin{subfigure}{6cm}
    \centering\includegraphics[width=5cm]{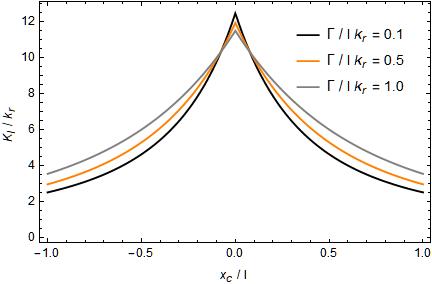}
    \caption{$x_{0}/l=0$, $\delta-$sink}
    \end{subfigure}
    
    \begin{subfigure}{6cm}
    \centering\includegraphics[width=5cm]{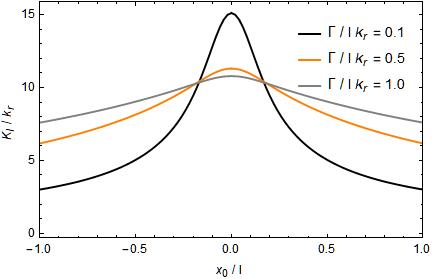}
    \caption{$x_{c}/l=0$, $\sigma/l=0.1$}
    \end{subfigure}
    \begin{subfigure}{6cm}
    \centering\includegraphics[width=5cm]{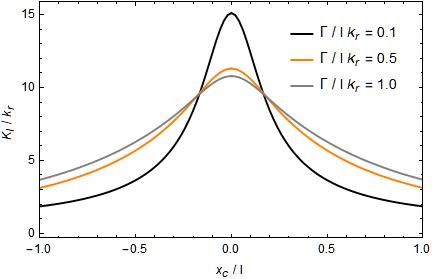}
    \caption{$x_{0}/l=0$, $\sigma/l=0.1$}
    \end{subfigure}
    
    \begin{subfigure}{6cm}
    \centering\includegraphics[width=5cm]{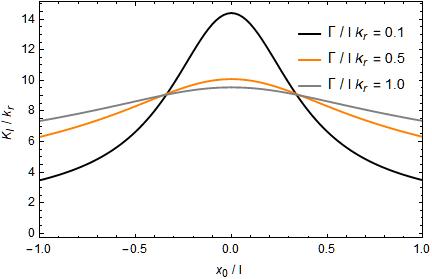}
    \caption{$x_{c}/l=0$, $\sigma/l=0.3$}
    \end{subfigure}
    \begin{subfigure}{6cm}
    \centering\includegraphics[width=5cm]{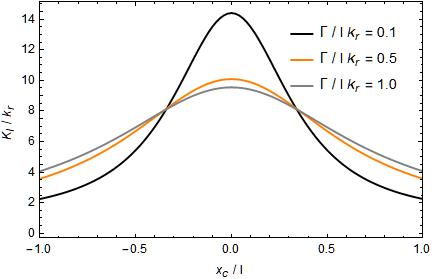}
    \caption{$x_{0}/l=0$, $\sigma/l=0.3$}
    \end{subfigure}
   \caption{\small{\emph{Average rate $K_{I}$ (in units of $k_{r}$) for different value of the slop $\Gamma$ (in units of $l k_{r}$) of the PEC with keeping the value of $l k_{r}/k_{c}$ to 0.05. (a) and (b) correspond to $\delta-$function sink, and the rest of all figures correspond to the Gaussian sink function. All the left panels in the above represent the case when the sink position $x_{c}$ is at the center (origin) of PEC, and the particle position $x_{0}$ is moving from left to right direction. All the right panels correspond to the opposite situation. An interesting physical phenomenon is observed in these figures. Near the center particle's diffusion motion leads over the drift and gives the highest rate, whereas away from the center, drift motion increases the rate irrespective of the relative position of the particle and the sink. The introduction of the sink of finite width enhances the distance over which diffusive influenced rate is obtained.}}}
\end{figure}

\begin{figure}[hbt!]
    \begin{subfigure}{6cm}
    \centering\includegraphics[width=5cm]{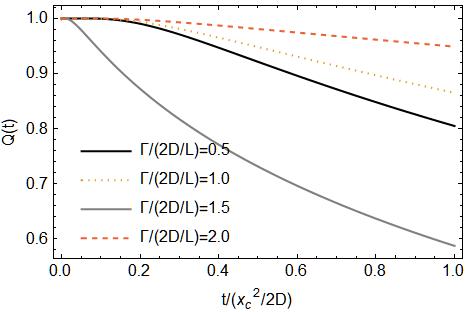}
    \caption{}
    \end{subfigure}
    \begin{subfigure}{6cm}
    \centering\includegraphics[width=5cm]{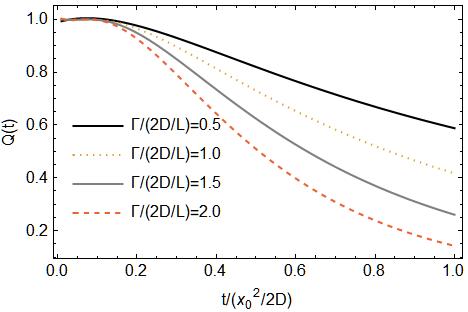}
    \caption{}
    \end{subfigure}

    \begin{subfigure}{6cm}
    \centering\includegraphics[width=5cm]{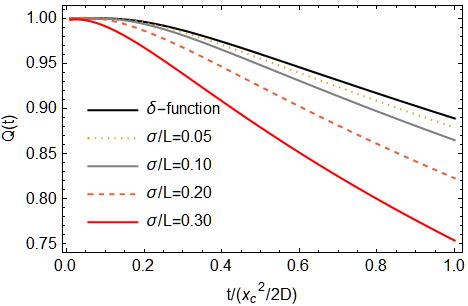}
    \caption{}
    \end{subfigure}
    \begin{subfigure}{6cm}
    \centering\includegraphics[width=5cm]{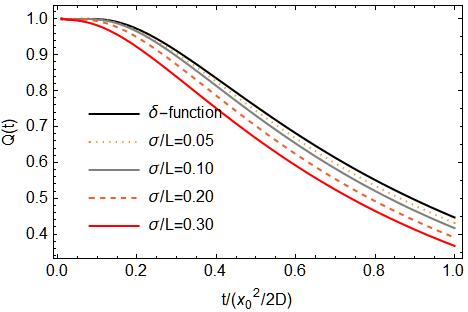}
    \caption{}
    \end{subfigure}
   \caption{\small{\emph{Survival probability Q(t) vs. time t (in units of $L^2/2D$) for the variation of strength and width of the trap function. (a) and (c), when $x_{0}=0$ and $x_{c}=L$. (b) and (d), when $x_{0}=L$ and $x_{c}=0$. With the variation of $\Gamma$ (in units of 2D/L) for the finite width $\sigma/L$=.1, we get nonmonotonic characteristics in (a) and monotonic nature in (b). A simple monotonic decay of $Q(t)$ is observed in both the figures (c) and (d) when $\Gamma/(2D/L)$ is set to 1. In the same figures, we plotted  Q(t) for the localized Dirac delta trap function ($\delta-$ function) for the better comparison. In every case, larger width corresponds to faster decay, and the value of the strength $k_{c}$ (in units of 2D/L) is set to 4.}}}
\end{figure}

\subsection{Quantum yield calculation in doped molecular crystal}  
In this section, we describe our method and calculate the quantum yield in doped molecular crystal \citep{wolf1968energy, powell1975singlet,kenkre1982master} and photosynthesis system \citep{clayton1980photosynthesis} based on this $"V"$ shaped potential with a general Gaussian sink. For the qualitative and quantitative study of exciton motion in an organic crystal (e.g., anthracene), guest molecules (e.g., tetracene) are brought up in small concentrations in the host crystal. Frenkel excitation is happened due to sinning light, and excitons are traveling through the host molecule. Finally they got trapped by guest molecules (considered to be a trap or a sink). The analogous processes happen in photosynthetic systems, where traps are the reaction centers, and the excitation gets captured by these traps and used for the farther process of making sugar. \citep{clayton1980photosynthesis}.\\
Host and guest molecules have different electronic excitation as a result of a different finite lifetime, and eventually, they decay radiatively. So one can easily track down the amount of the excitations placed originally into the host by illumination have been caught by, and emerged from, the guest. The ratio of the number eventually appearing from the host to the number placed initially in the host is known as the (host) yield and often denoted by $\phi$. If there are no nonradiative processes, then the energy transfer rate often expressed by $(1-\phi)/\phi$ and corresponding guest yield would be $1-\phi$. We calculate these quantities in the presence of the Gaussian trap in representative 1-dimensional molecular crystal. We define the lifetime of the excitation as $\tau$, and the relevant stochastic equation becomes
\begin{equation}
     \frac{\partial P(x,t|x_{0})}{\partial t}=[\mathscr{L}- \frac{k_{c}}{\sqrt{2 \pi \sigma^2}} e^{-\frac{(x-x_{c})^2}{2 \sigma^2}}-\frac{1}{\tau}]P(x,t|x_{0}),
\end{equation}
where $x_{c}$ is centre of the trap and $\sigma$ refers to width of the Gaussian trap. The expression of $\phi$ in terms of Greens function propagator is
\begin{equation}
    \phi=[1-\frac{k_{c}J_{1}(x_{0},\frac{1}{\tau})}{1+k_{c}J_{2}(x_{0},\frac{1}{\tau})/J_{1}(x_{0},\frac{1}{\tau})}]_{s=0}.
\end{equation}
We consider that the trap center is at $x_{c} = -L/2$ and calculate the
observable $\phi$ for the localized initial condition that initially the
excitation is at $x_{0} = L/2$.
\begin{figure}[hbt!]
\begin{subfigure}{6cm}
    \centering
    \includegraphics[width=5cm]{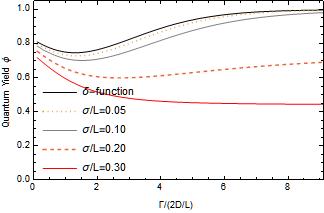}
    \caption{$\tau/(L^2/2D)=1$}
\end{subfigure}    
\begin{subfigure}{6cm}
    \centering \includegraphics[width=5cm]{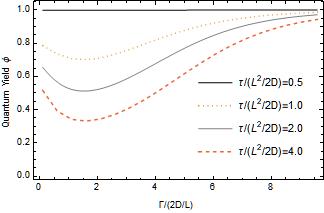}
    \caption{$\sigma/L=0.10$}
\end{subfigure}    
 \caption{\small{\emph{The quantum yield vs. the potential strength $\Gamma$ (in units of 2D/L) of the Frenkel excitation when $x_{0}$ and $x_{c}$ are separated by a distance L symmetrically around the origin. The capture parameter, $k_{c}$, is set to 4 (in units of 2D/L). (a) The quantum yield $\phi$ for three different values of the trap width $\sigma$ (in units of L) and also it is compared with the $\delta$ function trap. Clearly, a monotonic trend is observed with the increase in $\sigma$. (b) Nonmonotonic dependence of the quantum yield for four different values of the lifetime $\tau$ (in units of $L^2/2D$). Nonmonotonic effects (see text) are clear from the minimum in the quantum yield located at $\Gamma \approx 2D/L.$}}}
\end{figure}
\begin{figure}[hbt!]
\begin{subfigure}{6cm}
    \centering
    \includegraphics[width=5cm]{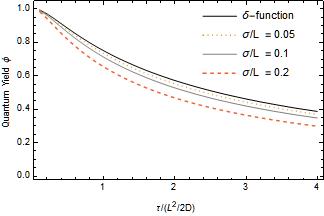}
    \caption{$\Gamma/(2D/L)=1$}
\end{subfigure}    
\begin{subfigure}{6cm}
    \centering \includegraphics[width=5cm]{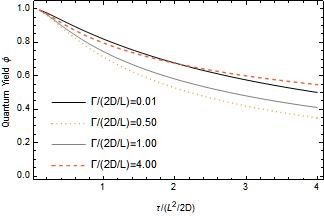}
    \caption{$\sigma/L=0.10$}
\end{subfigure}    
 \caption{\small{\emph{The quantum yield vs. the range of lifetime $\tau$ (in units of $L^2/2D$) of the Frenkel excitation when $x_{0}$ and $x_{c}$ are separated by a distance L symmetrically around the origin. The capture parameter, $k_{c}$, is set to 4 (in units of 2D/L). (a) The quantum yield $\phi$ for three different values of the trap width $\sigma$ (in units of L) and also it is compared with the $\delta$ function trap. Clearly, no nonmonotonic trend is observed with the increase in $\sigma$. (b) Nonmonotonic dependence of the quantum yield for four different values of the $\Gamma$ (in units of $2D/L$).The quantum yield vs. the range of lifetime $\tau$ (in units of $L^2/2D$) of the Frenkel excitation when $x_{0}$ and $x_{c}$ are separated by a distance L symmetrically around the origin. The capture parameter, $k_{c}$, is set to 4 (in units of 2D/L). (a) The quantum yield $\phi$ for three different values of the trap width $\sigma$ (in units of L) and also it is compared with the $\delta$ function trap. Clearly, no nonmonotonic trend is observed with the increase in $\sigma$. (b) Nonmonotonic dependence of the quantum yield for four different values of the $\Gamma$ (in units of $2D/L$).}}}
\end{figure}
Now we focus on the variation of Quantum yield $\phi$ on the strength $\Gamma$(in units of 2D/L) for the range of lifetime of the excitation and the width of the delocalized Gaussian trap. The dependency of $\phi$ on different parameters are depicted in Figures 5 and 6. Figure 5(a), where $\phi$ is plotted against width $\sigma$ (in units of L) of the trap function for four different values, and also calculated $\phi$ from the $\delta-$function trap is shown in the same figure. Here the value of $\tau/(L^2/2D)$ is assumed to be 1. Although characteristics of $\phi$ for $\delta-$function as well as for the small value of width are nonmonotonic but it loses its nonmonotonicity with the increase of the width. Figure 5(b), where curves are corresponding to different range of excitation lifetime $\tau$(in units of $ L^2/2D$), and the value of $\sigma/L$ is assumed to be 0.10. In this case, we do not get any different result due to the introduction of the Gaussian trap, and rather we get the same nonmonotonic effect as in the case of $\delta-$function trap.\\
Our concern is to explore the dependency of $\phi$ on the time domain, which can be realized partially by plotting $\phi$ against the lifetime $\tau$ of the excitation in Figure 6. Figure 6(a) shows the shape of $\phi$ for different width of the trap function, and the value of confinement strength $\Gamma$ is set to 1. Simple monotonic decay of $\phi$ is observed with $\tau$ irrespective of the width of the trap function. Figure 6(b), where different curves corresponding to different confinement strength $\Gamma$ with a fixed value of the trap function. We get the remarkable nonmonotonic behavior of $\phi$  against $\Gamma$ for the finite width of the trap function. The same characteristics is noticed before in \citep{chase2016analysis} for case of the $\delta-$function trap.
\subsection{Transfer rate calculation in different processes}
\begin{figure}[hbt!]
\begin{subfigure}{6cm}
    \centering
    \includegraphics[width=5cm]{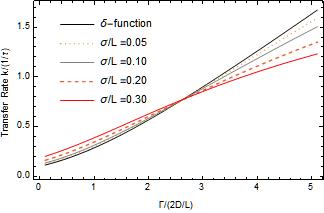}
    \caption{$x_{c}/L=0$}
\end{subfigure}    
\begin{subfigure}{6cm}
    \centering 
    \includegraphics[width=5cm]{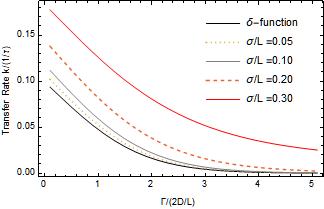}
    \caption{$x_{0}/L=0$}
\end{subfigure}\\
\begin{subfigure}{6cm}
    \centering
    \includegraphics[width=5cm]{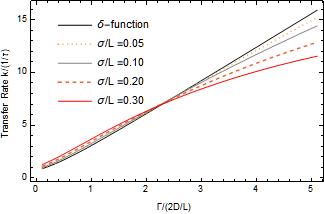}
    \caption{$x_{c}/L=0$}
\end{subfigure}    
\begin{subfigure}{6cm}
    \centering 
    \includegraphics[width=5cm]{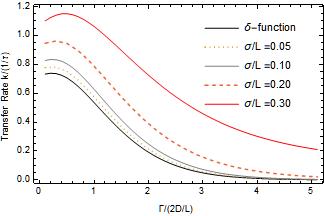}
    \caption{$x_{0}/L=0$}
\end{subfigure}
 \caption{\small{\emph{Transfer rate k (in units of $1/\tau$) vs. strength $\Gamma$ (in units of $2D/L$) of the potential with the increase in width $\sigma$ for the one-sided relative position of the particle and the trap on the PEC. (a) $\&$ (c), when the particle is at the uphill location, and the trap is at the center. (b) $\&$ (d), for the case when trap and the particle exchange their position i.e., the particle is at the center, and the trap is placed at the uphill location. Upper two figures(a) $\&$ (b) correspond to short probing time whereas lower two figures (c) $\&$ (d) are associated with relatively long probing time, and the probing time $\tau$ is 0.5 and 4 respectively. Trap strength $k_{c}$ is assumed to be 4 for all calculation.}}} 
\end{figure}
Apart from calculating survival probability (since it is not feasible to get the inverse Laplace transform analytically for every system) in the Laplace domain description, there should be a method of analysis and simplify the calculation without doing of numerical inversion of the Laplace transform. The finding of the transfer rate is one of the examples of those type for the kind of system described. Even we can extend this model to various systems depending on how we consider the parameter $\tau$ in the above equation. In the context of ecology, if the moving particle is assumed to be an animal, then the term $\tau$ could be its lifetime. Here the finiteness of $\tau$ is happened by natural causes or predators. In the previous application, the moving particle is an exciton, and its finite lifetime often corresponds to the radiative(as it turns into a photon) or sometimes to nonradiative (as it disappears in other way) depending upon the how they decay. If the physics of the system does not have a particle of a finite lifetime, $\tau$ in this present analysis would be referred to as a simple probe time. In the following, we calculate the transfer rate from the expression \citep{spendier2013reaction}.
\begin{equation}
    k=[\frac{1}{\hat{Q}(s)}-s]_{s=1/\tau}
\end{equation}
For our $"V"$ potential with the finite width of a delocalized trap function, the explicit form of the transfer rate constant would be
\begin{equation}
    k=\frac{1}{\tau}[\frac{J_{1}(x_{0},\frac{1}{\tau})}{\frac{1}{k_{c}}+J_{2}(x_{0},\frac{1}{\tau})/J_{1}(x_{0},\frac{1}{\tau})-J_{1}(x_{0},\frac{1}{\tau})}].
\end{equation}
To explore the transfer rate, we employ the above equation for the Gaussian trap function. We calculate the rate for the one-sided placement of the trap and initial distribution on the piecewise linear attractive potential and the difference in distance between the center of the trap and the initial position is L. The dependency of the transfer rate on the confinement strength of the potential for different width of the trap function (including Dirac delta trap) are plotted graphically in Figures 7(a), (b), (c), and (d). Figure 7(a), where k is plotted against $\Gamma$ (in units of 2D/L) for four different values of the width $\sigma$ (in units of L) when the trap function is placed at the center of the PEC and initial distribution is on either side (uphill location) of the PEC. With this situation, a crossover characteristic of k is found with respect to width $\sigma$ of the trap function. This nonmonotonic effect is quite surprising to us and intuitively nonfeasible. Figure 7(b) corresponds to interchange placement of the trap and initial distribution on the PEC. Here we observe a simple decrease in the value of k with $\Gamma$ and a higher value of $\sigma$ results high transfer rate. In these two figures, the probe time $\tau$ (lifetime in Frenkel excitation in photosynthesis) is chosen to be in the short-range, we consider the value of $\tau$ (in units of $L^2/2D$) as 0.5. During the calculation in this section, the strength of the trap function $k_{c}$ is 4. For the higher probing time $\tau$, the calculated value of k, keeping all the rest of the parameter value the same, is shown in Figure 7(c) and (d). We fix the value of $\tau$ to 4. In Figure 7(d), we see a small increment of k in the small range of $\Gamma$ and then decays to zero. These characteristics are well known and demonstrated before in case of harmonic potential in ref. \citep{spendier2013reaction}.

\section{Concluding Remarks}
The purpose of the paper has been to provide a mathematical method to solve the Smoluchowski equation of a particle moving under diffusion motion encountered by a  piecewise linear potential as well as Gaussian sink or trap on the same PEC. Our main finding is Eq. (17). Eq. (21) is found from considering first term approximation to the series solution of the quantity $\hat{R}(x,s)$. Importantly the solution in these expression is general for the arbitrary shape of the sink function as well as potential energy curve. 
Only the knowledge of the Greens function for the absence of any sink is required to get the solution in the Laplace domain. We use this expression throughout the next sections for different physical entities. For the validation of the expression, we often compare the calculated value corresponding to Gaussian sink with the one calculated from the exact expression for the $\delta-$sink function. Figures 1, 2, $\&$ 3 describe the observable average rate with the variation of different dependable parameters. The main findings in this section are this rate depends upon the width of the sink function non monotonically when the particle is near to the sink. When they are far from each other, the rate has become independent of the shape of the sink function. With the variation of the confinement strength of the potential, both monotonic and nonmonotonic dependence of the rate is observed in Figure 3. Things to notice that the distance over which the nonmonotonicity $l_{nmt}$ exist increases with the width of the sink function. Numerically calculated the survival probability of the diffusion dynamics is shown in Figure 4. Here we see expected nonmonotonic dependency of the probability with the variation of the strength of the PEC for the uphill location of the sink function with respect to the particle position. But it does not happen in the opposite situation. An enhancement of the decay of the survival probability is happened with the increase of the width of the sink irrespective of two different types of situations. We have calculated quantum yield by using Eq. (24) for various systems like doped molecular crystal and photosynthesis processes involving Frenkel excitation. The dependency of this observable on strength and the lifetime is depicted in Figures 5 $\&$ 6. With the variation of the strength, quantum yield changes non monotonically for a smaller value of the width, and it changes monotonically for a farther increase of the width. Finally, we calculate the transfer for rate by using the expression in Eq. (26). Here we see initially transfer rate has a higher value corresponding to a higher value of the sink width for the variation of the strength when the sink is at the center of the PEC and particle is at the uphill location. But it posses lower value corresponds to the higher value of the width for a farther increase in the strength. This characteristic is shown in Figure 7 (a) $\&$ (c). For the opposite situation of the particle and the sink position, the transfer rate increases with the strength and then decays monotonically irrespective of the different value of the width. Overall our study uncovered many unknown features as well as some existence features of different observable for the introduction of general Gaussian sink. Solving the equation for the common harmonic potential energy curve with the general Gaussian sink could be an interesting future work over this model.
\section*{Acknowledgement}
One of the author (C.S.) would like to thank the Indian Institute of Technology Mandi for Half-Time  Research Assistantship (HTRA) fellowship.
\bibliography{rf}
\bibliographystyle{unsrt}

\end{document}